\documentclass[aip,apl,amsmath,amssymb,reprint]{revtex4-1}
\usepackage{graphicx}
\usepackage{dcolumn}
\usepackage{bm}
\usepackage{mathrsfs} 
\usepackage{color}

\usepackage{lineno}

\usepackage{siunitx}

\newcommand{\Xv}{$\langle 100 \rangle$}

\usepackage{xcolor}

\begin{document}

\preprint{AIP/123-QED}

\title[Charge Transport in Silicon]{High-field Spatial Imaging of Charge Transport in Silicon at Low Temperature}

\affiliation{Department of Physics, Stanford University, Stanford, CA 94305 USA}
\affiliation{SLAC National Accelerator Laboratory/Kavli Institute for Particle Astrophysics and Cosmology, 2575 Sand Hill Road, Menlo Park, CA 94025 USA}
\affiliation{Fermi National Accelerator Laboratory, Center for Particle Astrophysics, Batavia, IL 60510 USA}
\affiliation{Kavli Institute for Cosmological Physics, University of Chicago, Chicago, IL 60637, USA}
\affiliation{Department of Physics, Santa Clara University, Santa Clara, CA 95053 USA}
\affiliation{Department of Physics, University of Illinois at Urbana-Champaign, Champaign, IL 61820 USA}
\affiliation{Department of Physics, San Diego State University, San Diego, CA 92182 USA}

\author{C.~Stanford} \email{cstan4d@stanford.edu}\affiliation{Department of Physics, Stanford University, Stanford, CA 94305 USA}

\author{R.A.~Moffatt} \affiliation{Department of Physics, Stanford University, Stanford, CA 94305 USA}

\author{N.A.~Kurinsky} \affiliation{Fermi National Accelerator Laboratory, Center for Particle Astrophysics, Batavia, IL 60510 USA} \affiliation{Kavli Institute for Cosmological Physics, University of Chicago, Chicago, IL 60637, USA}

\author{P.L.~Brink} \affiliation{SLAC National Accelerator Laboratory/Kavli Institute for Particle Astrophysics and Cosmology, 2575 Sand Hill Road, Menlo Park, CA 94025 USA}

\author{B.~Cabrera}\email{cabrera@stanford.edu} \affiliation{Department of Physics, Stanford University, Stanford, CA 94305 USA} \affiliation{SLAC National Accelerator Laboratory/Kavli Institute for Particle Astrophysics and Cosmology, 2575 Sand Hill Road, Menlo Park, CA 94025 USA}

\author{M.~Cherry} \affiliation{SLAC National Accelerator Laboratory/Kavli Institute for Particle Astrophysics and Cosmology, 2575 Sand Hill Road, Menlo Park, CA 94025 USA}

\author{F.~Insulla} \affiliation{Department of Physics, Stanford University, Stanford, CA 94305 USA}

\author{M.~Kelsey} \affiliation{SLAC National Accelerator Laboratory/Kavli Institute for Particle Astrophysics and Cosmology, 2575 Sand Hill Road, Menlo Park, CA 94025 USA}

\author{F.~Ponce} \affiliation{Department of Physics, Stanford University, Stanford, CA 94305 USA}

\author{K.~Sundqvist} \affiliation{Department of Physics, San Diego State University, San Diego, CA 92182 USA}

\author{S.~Yellin} \affiliation{Department of Physics, Stanford University, Stanford, CA 94305 USA}

\author{B.A.~Young} \affiliation{Department of Physics, Santa Clara University, Santa Clara, CA 95053 USA}

\date{\today}

\begin{abstract}
We present direct imaging measurements of charge transport across a 1\,cm$\times$1\,cm$\times$4\,mm-thick crystal of high purity silicon ($\sim$15 k$\Omega$-cm) at temperatures of 5\,K and 500\,mK. We use these data to measure the lateral diffusion of electrons and holes as a function of the electric field applied along the [111] crystal axis, and to verify our low-temperature Monte Carlo software. The range of field strengths in this paper exceed those used in the previous study~\cite{moffat2019} by a factor of 10, and now encompasses the region in which some recent silicon dark matter detectors operate~\cite{hvevRun1}.  We also report on a phenomenon of surface charge trapping which can reduce expected charge collection. 
\end{abstract}

\maketitle

\section{Introduction}

\begin{figure}[t]
    \centering
    \includegraphics[width=0.23\textwidth]{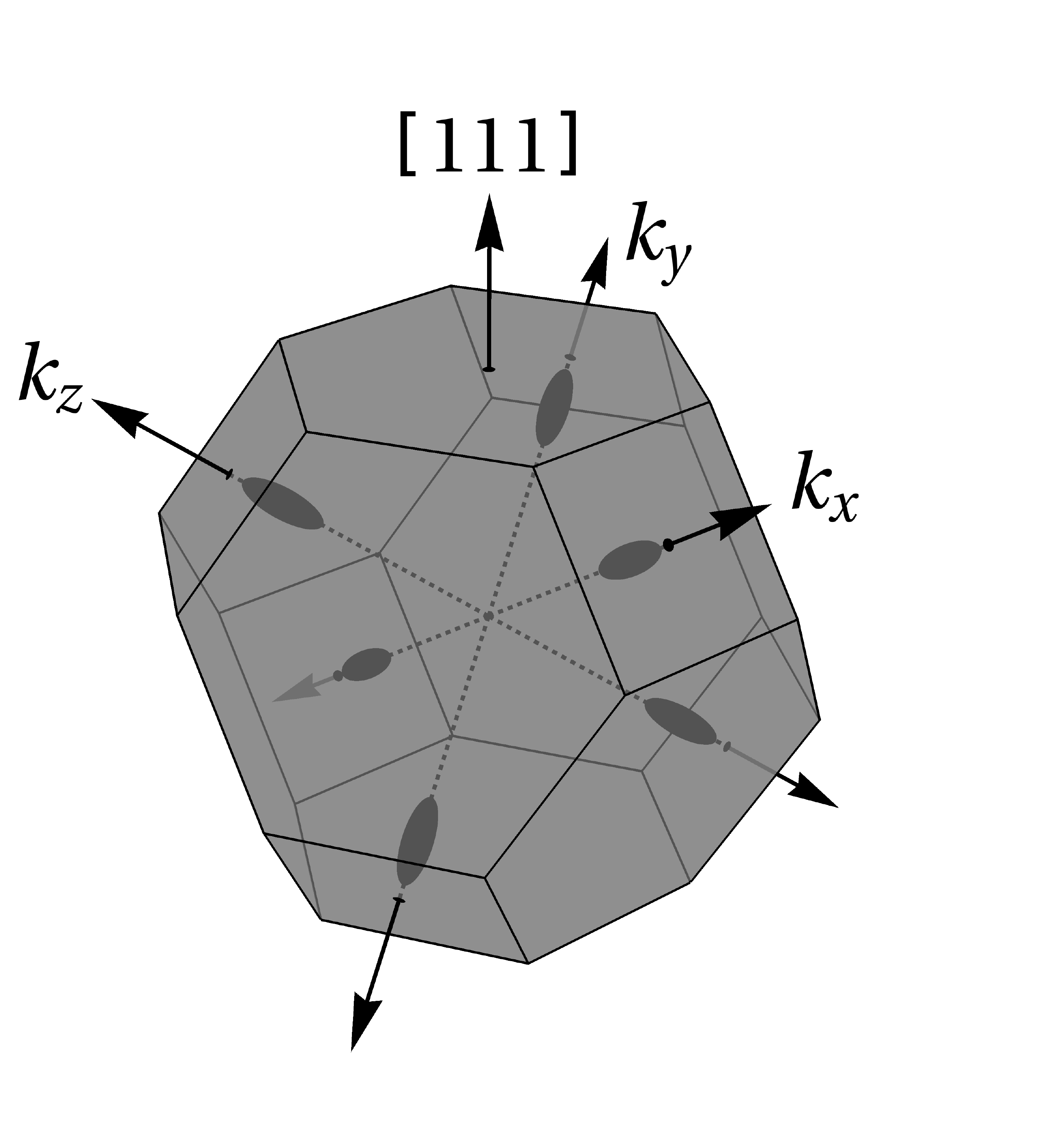}
    \includegraphics[width=0.23\textwidth]{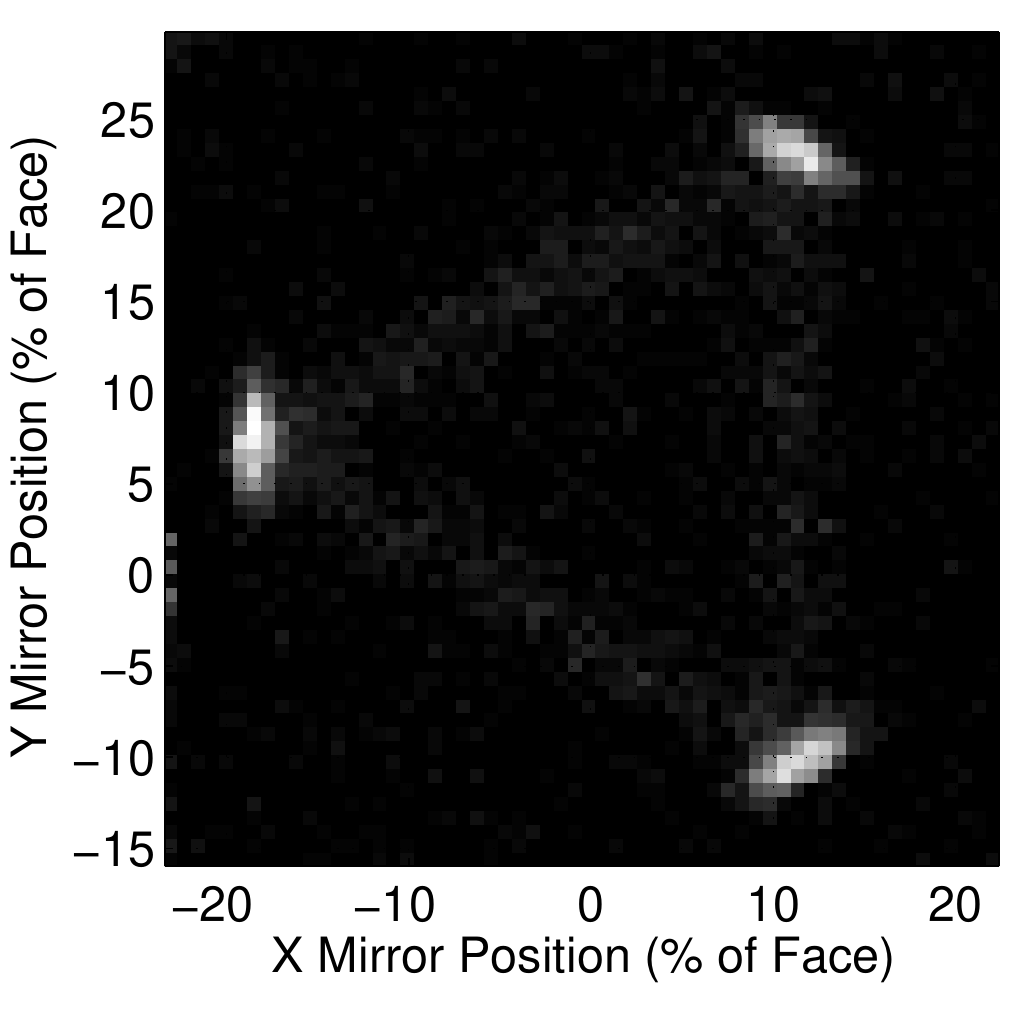}
    \includegraphics[width=0.48\textwidth]{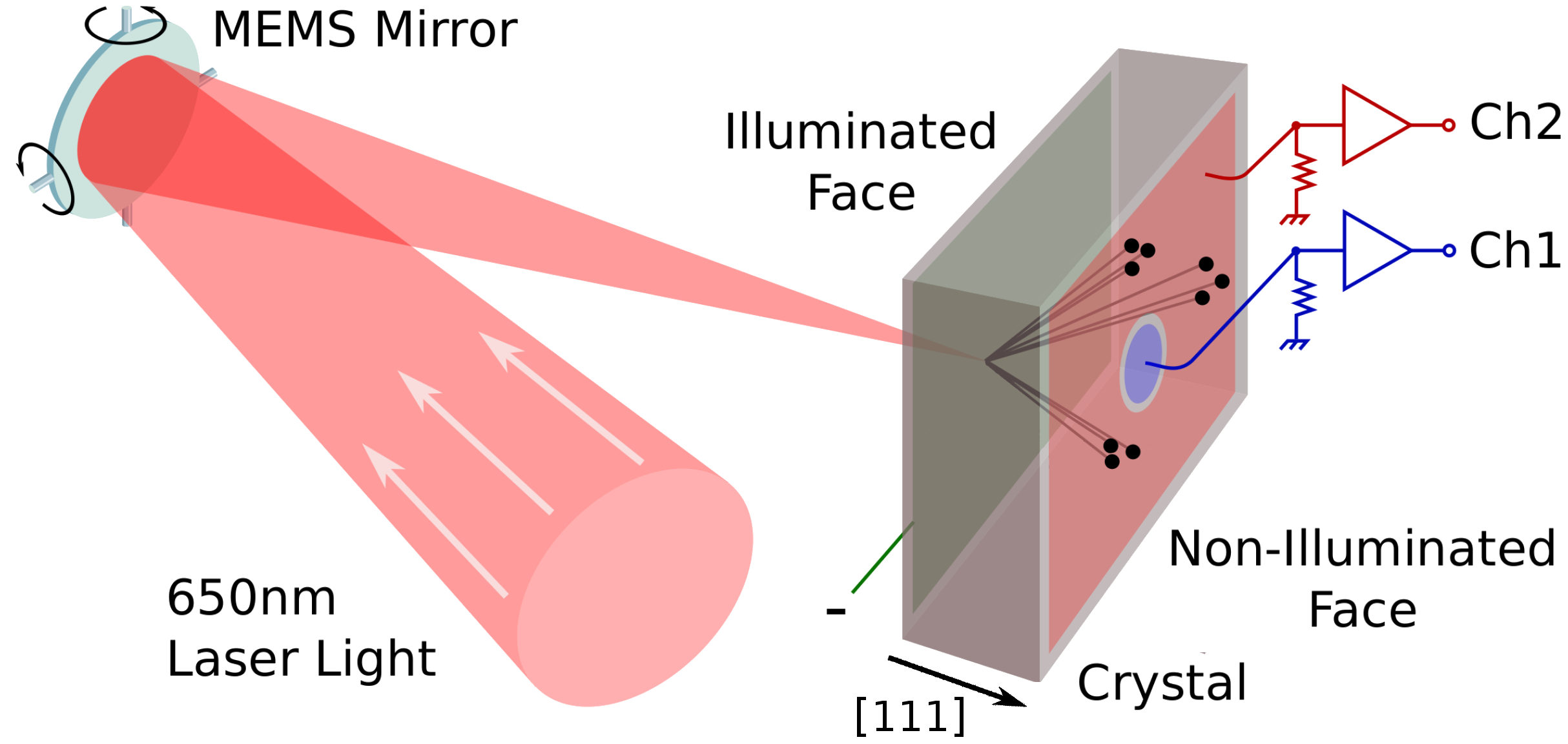}
    \caption{\textbf{Top left}: The first Brillouin zone for (111) Si. The electron energy minima, or valleys, are located near the edge of the zone in the $k_x$, $k_y$, and $k_z$ directions (black ovals). At low fields, electrons propagate along the valley axes and produce three distinct charge populations. 
    \textbf{Top right}:~A measurement of the charge pattern at low field (\SI{10}{\volt\per\centi\meter}), with the white spots indicating the regions where more charge was collected. \textbf{Bottom}:~The detector, which consists of a \SI{1}{\centi\meter}$\times$\SI{1}{\centi\meter}$\times$\SI{4}{\milli\meter} crystal of high purity silicon, and the laser mechanism, which uses a MEMS mirror to scan a focused beam of light pulses across the back face of the crystal in order to produce a 2D image. The crystal is biased via a parquet electrode on the illuminated face.}
    \label{fig:brillouin}
\end{figure}

Silicon (Si) is an indirect band gap semiconductor with electron energy minima (valleys) that are displaced from zero in momentum space along the \Xv\ crystal axes. The top left panel of Figure~\ref{fig:brillouin} shows the Brillouin zone for (111) Si, including the surfaces of constant energy about these minima. They appear oval in shape because they are highly anisotropic; electrons have effective masses of 0.98~$m_{e}$ and 0.19~$m_{e}$ in the momentum space directions parallel and perpendicular to the valley axis, respectively\cite{Jacoboni}. If a low electric field is applied to the crystal along the [111] direction, free electrons will tend to remain aligned with their individual valley axis rather than the applied field. Three distinct charge populations will then propagate through the crystal and reach the surface, as shown in the top right panel of Fig.~\ref{fig:brillouin}. 

As the electric field is increased, each drifting electron carries more energy. Energetic electrons are capable of emitting an optical phonon and transitioning between Brillouin zone minima\cite{Jacoboni,ridley1999,SundqvistThesis,canali}; this is called an intervalley scattering event. At higher field strengths, this scattering becomes so frequent that the electrons effectively propagate in the mean valley direction, in this case along [111] (which we label the $z$-direction). At the same time, the propagating electrons move with an $xy$-diffusion consistent with a 2-D random walk. We directly imaged this phenomenon in our previous paper~\cite{moffat2019} up to field strengths of \SI{50}{\volt\per\centi\meter}.

In contrast, holes in Si have their energy band minima at $k=0$ in momentum space, and therefore propagate along the applied field even at low strengths. The resulting spatial distribution of charges arriving at the Si surface is still not perfectly Gaussian, however, due to the inherently anisotropic dispersion relation for holes, as detailed in other papers~\cite{DKK,kane,cardona} and first imaged in \citet{moffat2019}.

In this paper, we present a direct measurement of the $xy$-diffusion of electrons and holes in (111) Si under electric field strengths up to \SI{500}{\volt\per\centi\meter}, a factor of 10 increase over our previous paper~\cite{moffat2019}. With these field strengths, we are able to achieve the same conditions as those used with our SuperCDMS High Voltage eV-scale (HVeV) dark matter detectors~\cite{hvevRun1}, so the diffusion measurements made here will allow us to refine the charge transport models used in the SuperCDMS detector Monte Carlo. In particular, we find that when an ionization site occurs near the crystal's side wall, the $xy$-diffusion can result in some of the charges hitting the wall and stopping before reaching the charge collection electrodes. Such fractional charge collection events are a significant background in the HVeV detectors~\cite{hvevRun1}, so in order to perform a background subtraction it is critical to have a proper model of the $xy$-diffusion in these devices. 

This paper also reports on the phenomenon of charge buildup on the surface of Si crystals, as it was discovered to be a relevant factor when modelling the diffusion of both electrons and holes. We observed that as propagating charges approached a cluster of like-charges trapped near the (111) surface electrodes, their paths were deflected outward (away from the $z$-direction), resulting in greater $xy$ diffusion than expected.

\section{Experimental Setup}\label{sec:exp}

The crystal under test was cut from a \SI{4}{\milli\meter}-thick wafer of undoped ultra-high-purity float-zone silicon ($\sim$15 k$\Omega$-cm). The residual impurity was measured to be p-type with a concentration of $10^{12}$\,cm$^{-3}$. The front (illuminated) and back (non-illuminated) faces of the crystal are \SI{1}{\centi\meter}$\times$\SI{1}{\centi\meter}, and lie in the (111) plane. The front face is patterned with an aluminum-tungsten mesh electrode, with 20\% coverage~\cite{electrode}. The back face features a small, inner electrode in the center of the face, circular in shape with a diameter of \SI{160}{\micro\meter}, separated by a \SI{10}{\micro\meter} gap from the outer electrode, which covers the rest of the face. These electrodes are separately contacted, but are held at the same voltage to provide a near-uniform field throughout the crystalline substrate. 

Free electrons and holes are produced at the front face of the crystal by exposing it to pulses of 650\,nm laser light, focused to a \SI{60}{\micro\meter} diameter spot. An exposure is typically a ``pulse train" of twenty \SI{200}{\micro\watt}, \SI{50}{\nano\second} pulses, with a 1\,ms delay between each pulse. Either the electrons or holes (depending on the sign of the bias voltage) freed by a light pulse will quickly propagate through the crystal and produce a voltage pulse in the electrodes on the back face of the detector. The data acquisition system records a \SI{20}{\milli\second} voltage trace for both the inner and outer electrodes. In post-processing, the 20 pulses contained in each \SI{20}{\milli\second} trace are averaged to obtain a single value for the pulse amplitude for each electrode.

In order to reduce scattering of propagating charges by background phonons, the crystal is cooled in a He-3 cryostat with the capability of reaching temperatures down to 500\,mK.

The position of the laser's focal point on the front face of the detector is controlled by a Micro-Electromechanical Systems (MEMS) mirror~\cite{mirrorcle}. 
During each imaging measurement the mirror is moved in such a way as to make the laser perform a raster scan across the front face of the crystal, pausing briefly at each position to enable periodic neutralization of the detector. The precise neutralization procedure involves electrically grounding both faces of the detector while illuminating the crystal with diffuse \SI{650}{\nano\meter} light for 1 second, before reapplying the voltage bias. This process ensures that the charge trapping sites in the crystal are neutralized before the pulse train is sent through.

At every point in the raster scan, the small inner electrode is sampling a different part of the charge pattern that arrives at the back face. So, by plotting the average amplitude of the inner electrode as a function of laser spot position, we obtain a 2-D image of the charge pattern.

\section{Charge Patterns}\label{sec:patterns}

\begin{figure}[t]
    \centering
	\includegraphics[width=0.48\textwidth]{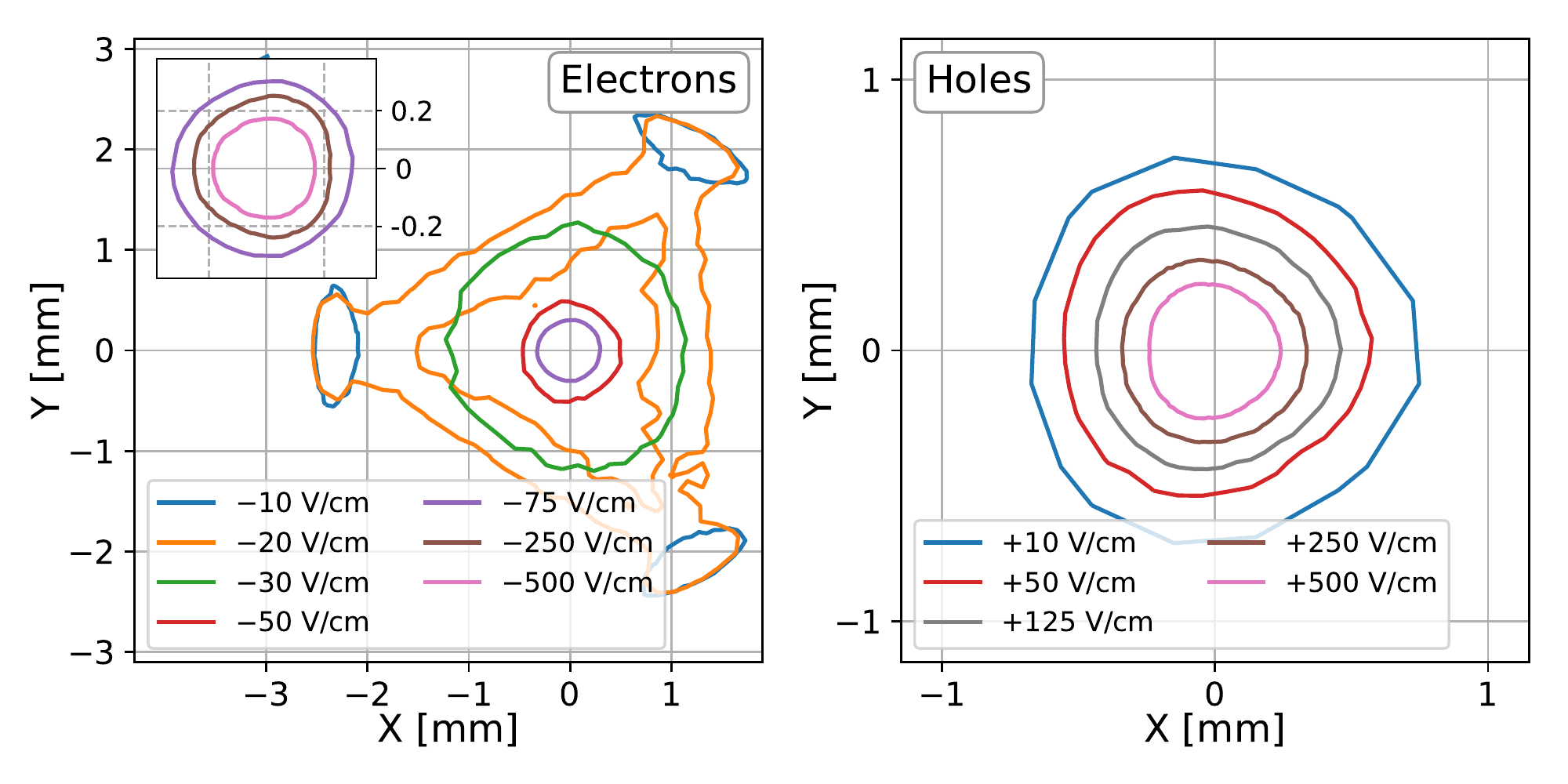}
    \caption{The 50\% contours for the charge charge density patterns for electrons (left) and holes (right) for electric field strengths up to \SI{500}{\volt/\centi\meter}. Note differences of scale between the figures. The electron patterns are much larger at low voltages due to the low rates of intervalley scattering. At high field strengths they scatter frequently and travel along the average valley direction.}
\label{fig:HV_contours}
\end{figure}

The first measurements made in this new study yielded full, 2-D charge pattern images for both electrons and holes over a range of field strengths from \SI{10}{\volt\per\centi\meter} to 500\,V/cm. The images were then reduced to their 50\% contours and overlaid with each other, as shown in Figure~\ref{fig:HV_contours}. The region enclosed in each colored line thus contains 50\% of the charge pattern for its corresponding voltage.

Figure~\ref{fig:HV_contours} (left) shows that at \SI{-10}{\volt/\centi\meter} (blue), the electrons largely remain in their original valleys and thus produce the expected ``tri-lobe" shape. At \SI{-20}{\volt/\centi\meter} (orange), the electron mean free path between intervalley scatters is on the order of the crystal's \SI{4}{\milli\meter} thickness. Electrons scattering only once will end up somewhere between the three lobes, which leads to a line appearing between the distinct tri-lobe populations. 
At \SI{-30}{\volt/\centi\meter} (green), the electron mean free path is just a fraction of the crystal dimensions, so most of the propagating charges scatter multiple times on their way to the electrodes. In this case, the larger applied electric field also forces the propagating charges closer to the [111] symmetry axis, resulting in the corresponding charge density patterns appearing towards the middle of the detector face.
At even stronger negative bias fields the detected charge density patterns remain roughly Gaussian in shape but with decreasing width for greater field strength and thus decreased electron mean free path.

The evolution of the hole pattern is more straightforward, as the holes propagate along the electric field even at low field strengths, producing a distorted Gaussian shape that shrinks with increasing voltage (see Figure~\ref{fig:HV_contours} (right)).

After observing the relatively small widths of the charge density patterns at the higher field strengths, we became concerned that self-repulsion of the propagating charges might be a relevant factor in the diffusion process. 
When we decreased the laser power to see if that would change the width of the observed charge density pattern we found that it did. This result launched us into an even more detailed investigation of the electron and hole dynamics present in our detector.

\section{Charge Trapping}\label{sec:trapping}
As mentioned in Section~\ref{sec:exp}, at each point in the raster scan, a train of twenty charge pulses is sent through the crystal. In our initial studies, the resultant voltage pulses were averaged to obtain a single value at each point in the scan. This methodology assumed that the crystal reverts to equilibrium during the 1\,ms between consecutive pulses. The basis for this assumption was that the transit time for the charges is on the order of hundreds of nanoseconds~\cite{canali}, and the recovery time of our amplifier circuit is on the order of \SI{100}{\micro\second}.

To test this assumption, we increased the power and duration of the laser pulses. We observed that even at high field strengths of \SI{100}{\volt/\centi\meter}, the patterns deviated significantly from the Gaussian spot obtained with lower laser intensities. Furthermore, we observed the charge pattern change over the course of the sequence of twenty pulses.

\begin{figure}[t]
    \centering
	\includegraphics[width=0.48\textwidth]{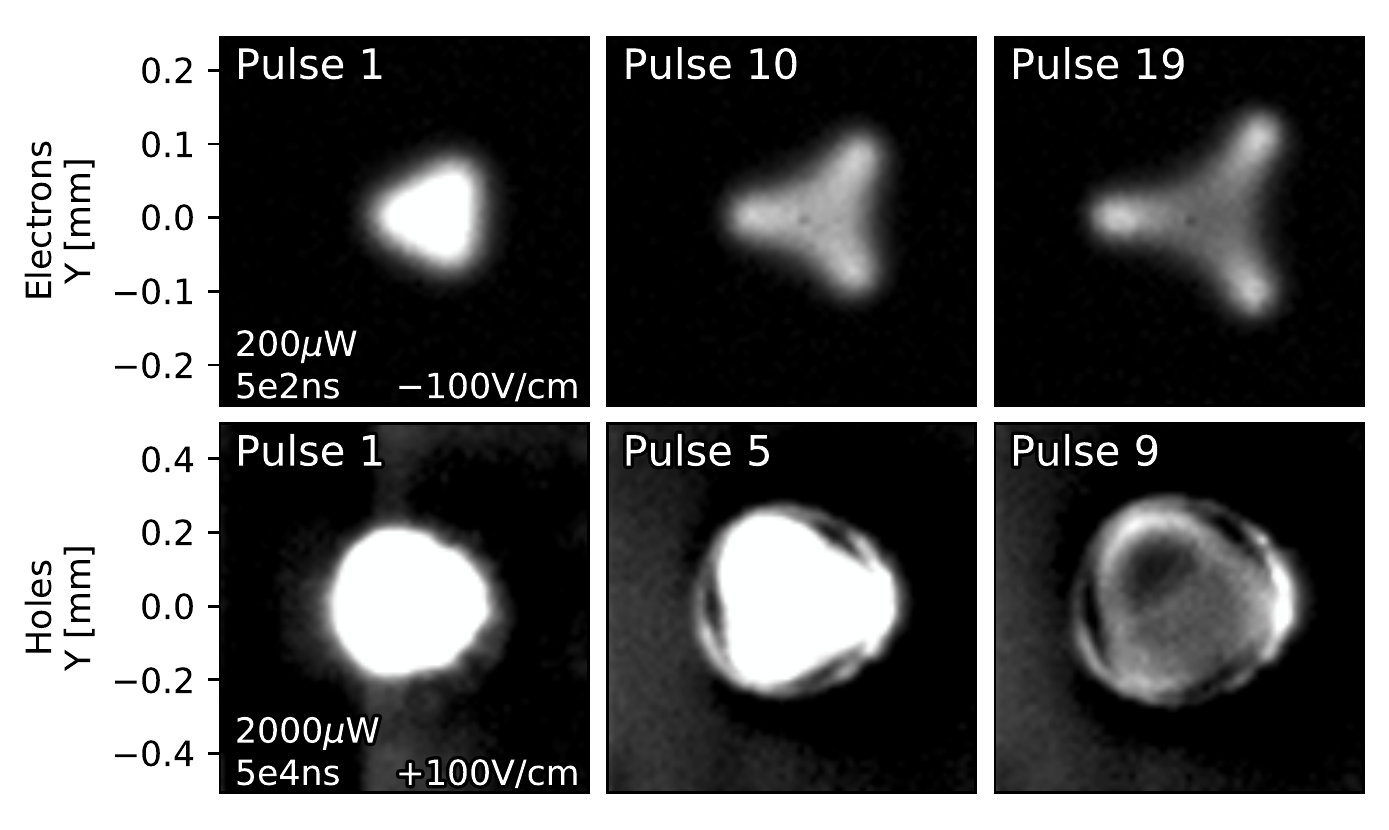}
    \caption{The evolution of the charge pattern as a function of pulse number for electrons (top row) and holes (bottom row). The white text in the bottom of the first figure in each row shows the power and width of the laser pulse and the bias applied to the crystal.}
\label{fig:surface_trapping_repulsion}
\end{figure}

The top row of Figure~\ref{fig:surface_trapping_repulsion} shows the evolution of the observed charge density pattern for electrons produced by \SI{200}{\micro\watt}, \SI{500}{\nano\second} wide laser pulses in a \SI{-100}{\volt\per\centi\meter} applied electric field. Electrons generated by the first laser pulse in each pulse series produced a fuzzy triangle-shaped pattern at the crystal surface (top-left panel). The pattern was different from the Gaussian-like spot seen at lower intensity (Figure~\ref{fig:HV_contours}). Pulses later in a single pulse sequence yielded increasingly distinct triangular patterns. Notably, these triangular shapes (see Figure~\ref{fig:surface_trapping_repulsion}, top-right panel) had the same orientation as the patterns observed previously at low electric field (Figure~\ref{fig:HV_contours}), where intervally scattering was infrequent.

The bottom row of Figure~\ref{fig:surface_trapping_repulsion} shows the evolution in the hole pattern as a function of pulse number. In contrast with the electron pattern, the hole collection pattern does not retain the same symmetry as the pulse number is increased; instead, we see a distinct ring appear outside the fuzzy triangular spot. 

\begin{figure}[t]
    \centering
	\includegraphics[width=0.49\textwidth]{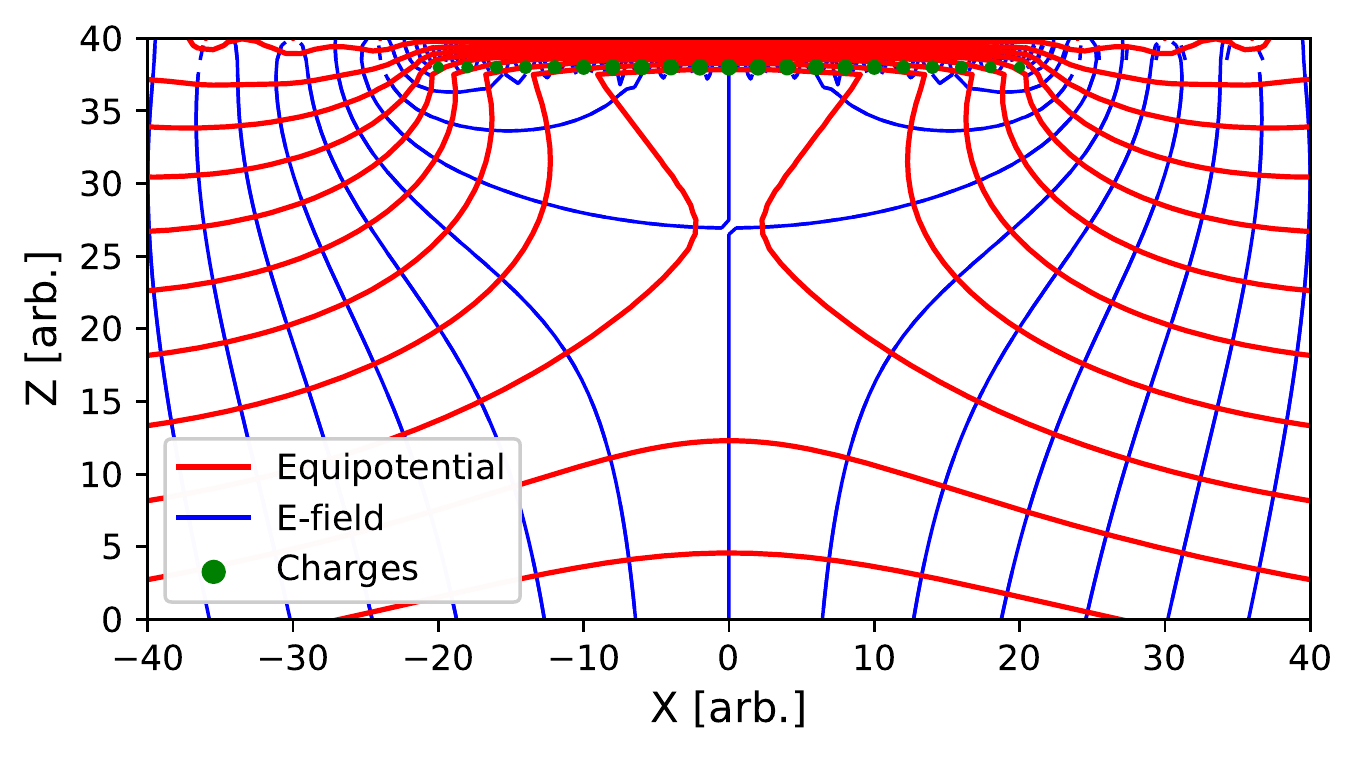}
    \caption{A 2-D toy model of the equipotential and field lines resulting from a series of charges stuck near one face of a parallel plate capacitor. $Z=40$ represents the back face of the crystal, with the charges placed at $Z=38$ stretching from $X=-20$ to $X=20$ with a Gaussian weighting. The front face of the crystal is at $Z=-40$ and is not shown. As charges travel upward along $X=0$, they will experience a potential minimum at $Z=27$ and move outward, producing a charge pattern that is brighter near edges than in its center.}
\label{fig:field_lines}
\end{figure}

Our best explanation for this behavior is that a fraction of the charges from each pulse are not collected by the electrodes and instead get stuck inside the crystal near the back (non-illuminated) face. The first few charges of the first pulse are unaffected, but later charges from the first pulse and charges from later pulses experience a distorted electric field from the stuck charges as they approach the electrodes. Figure~\ref{fig:field_lines} shows the electrostatic solution to a 2-D model of the proposed mechanism. Charges drifting upward along $X=0$ experience a saddle point in the electric field before reaching the surface. Some of these charges then drift horizontally until they reach an area of the surface with low charge density. This results in a charge pattern for the pulse that is more diffuse, with lower charge density at the center compared to the edges.

The model also indicates that this horizontal drift occurs in a region with relatively low field strength. The re-emergence of the tri-lobe structure in the electron pattern is therefore a result of the charges experiencing low rates of intervalley scattering during their horizontal drift. 

A low-field region inside the crystal would also explain the hole patterns. The band structure for holes has three valleys, two of which are degenerate at zero crystal momentum but anisotropic in position space. The third energy valley for holes (the ``split-off" band) is $\sim$44~meV higher in energy at zero momentum, but isotropic\cite{moffat2019,kane,ottaviani}. It appears that the holes are concentrating in this low-field region and filling the lowest energy valleys, forcing some charges into the split-off band. Then, the same horizontal drift phenomenon that occurs with the electrons occurs with the holes. The charges in the lower energy bands drift outward in $xy$ in an anisotropic way, forming a fuzzy triangle, while the charges in the split-off band drift outward isotropically, forming a distinct ring. We believe this to be the first direct imaging of the split-off band.




To reduce the image distortion from charge trapping for the remainder of the study, we used a 20\,dB fiber attenuator to significantly decrease the laser intensity. We then plotted the standard deviation ($\sigma$) of the charge pattern size as a function of pulse number (1-20). Noticing that the results for the initial pulses for each sequence indicated a $\sigma$ that increased linearly with pulse number, we fit the $\sigma$ for the first five pulses with a linear function and used the intercept for this fit in the final result, presented below.  

\section{Diffusion Results}\label{sec:results}

\begin{figure}[t]
\centering
\includegraphics[width=0.49\textwidth]{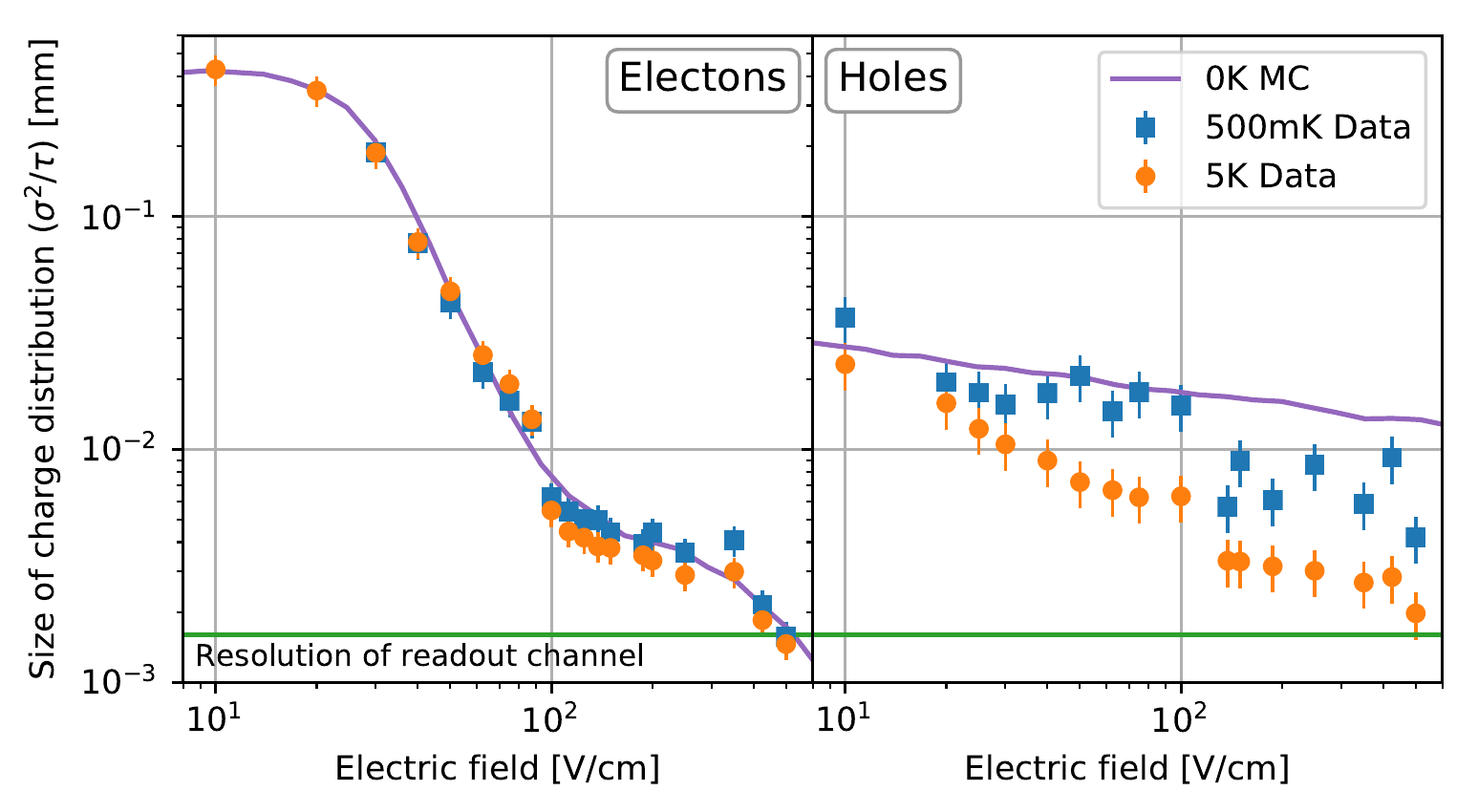}
\caption{The relative size of the charge pattern for electrons (left) and holes (right) as measured by the variance of the charge pattern ($\sigma^2$) normalized by the crystal thickness ($\tau$). The Monte Carlo (solid line) was fit to the electron data with a physical model involving 3 free parameters (see text). The Monte Carlo for the holes uses just one free parameter to model their scattering rate.}
\label{fig:angular_size}
\end{figure}

The final results of the study are shown in Figure~\ref{fig:angular_size}. In this plot, the size of the charge collection pattern, as measured by the variance ($\sigma^2$), is normalized by the crystal thickness ($\tau$) to provide a crystal-independent value for the lateral charge diffusion. For the low voltage electron data ($<$\SI{30}{\volt\per\centi\meter}), where the tri-lobe pattern is still present, $\sigma^2$ is a measurement of the variance of the distribution after being projected onto the $x$-axis. For the higher voltage electron data, and for the hole data, where the charge pattern is approximately Gaussian, $\sigma^2$ was obtained from Gaussian fit of a 1D scan through the center of the pattern.

Overlaid on each plot is the results of our Geant4 Condensed Matter Physics (G4CMP) simulation software, which we developed to extend the capabilities of Geant4 into the regime of low-temperature charge and phonon transport~\cite{brandt}. It includes the electron valleys in silicon and the intervalley scattering process. The scattering is implemented as a Poisson process with a rate function defined by
\begin{equation}
    R = \Gamma_I + \Gamma_{Ph} (|E|/\SI{}{\volt\per\centi\meter})^\alpha
\end{equation}
where $\Gamma_I$ and $\Gamma_{Ph}$ are, respectively, the rate constants associated with impurity scatters, which dominate at low voltage, and phonon-assisted transitions, which dominate at high voltage. A third constant, $\alpha$, is used to set the dependence of the phonon-assisted transitions on the electric field~($E$).

 The best-fit parameters for the data up to \SI{500}{\volt\per\centi\meter} are $\Gamma_I=\SI{1.5(5)e6}{\hertz}$, $\Gamma_{Ph}=\SI{1.5\pm0.3}{\hertz}$, and $\alpha=\SI{4.0\pm0.1}{}$.
 These values ($\Gamma_{Ph}$ in particular) differ from the results of our previous paper because the value for the phonon-assisted transition rate was not well constrained for the voltage range we were probing ($2$ to \SI{50}{\volt\per\centi\meter}).

\acknowledgements{This work was supported in part by the U.S. Department of Energy and by the National Science Foundation. This document was prepared by using resources of the Fermi National Accelerator Laboratory (Fermilab), a U.S. Department of Energy, Office of Science, HEP User Facility. Fermilab is managed by Fermi Research Alliance, LLC (FRA), acting under Contract No. DE-AC02-07CH11359. SLAC is operated under Contract No. DEAC02-76SF00515 with the U.S. Department of Energy.}
\bibliographystyle{aipnum4-1}
\bibliography{APL_Si_Transport}

\end{document}